\documentclass{appolb}
\usepackage{epsfig}
\usepackage{color}
\usepackage{cite}
\usepackage{bm}        
\usepackage{amssymb}   
\usepackage{hyphenat}
\usepackage[section]{placeins}
\newcommand{\rmd}{\mathrm{d}}

\begin{document}
\pagestyle{plain}
\newcount\eLiNe\eLiNe=\inputlineno\advance\eLiNe by -1
\title {Interplay between endogenous and exogenous fluctuations \\ in financial markets}
\author{V. Gontis\footnote{e-mail: vygintas@gontis.eu}
\address{Institute of Theoretical Physics and Astronomy \\ Vilnius University 
\\ 
Vilnius, LT 01108, Lithuania}}
\maketitle

\begin{abstract}
We address microscopic, agent based, and macroscopic, stochastic, modeling of the financial markets combining it with the exogenous noise. The interplay between the endogenous dynamics of agents and the exogenous noise is the primary mechanism responsible for the observed long-range dependence and statistical properties of high volatility return intervals. By exogenous noise we mean information flow or/and order flow fluctuations. Numerical results based on the proposed model  reveal that the exogenous fluctuations have to be considered as indispensable part of comprehensive modeling of the financial markets. 
\end{abstract}

\section{Introduction}
Statistical physics 
has been found useful dealing with the general concepts of complexity 
and its applications in finance \cite{Karsai2012NIH,Chakraborti2011RQUF1,Gabaix2009AR}. Financial markets are among the most interesting examples of such complex social systems where methods of statistical physics face extreme challenges \cite{Farmer2012EPJ}. Although the contemporary understanding of the nature of
microscopic market interactions is limited and ambiguous
\cite{Shiller2014AER,Kirman2014MD}, the advanced methods of
empirical data analysis and agent based modeling are very useful to gain greater insight into the market's complexity
\cite{Mantegna2000Cambridge,Bouchaud2004Cambridge,Sornette2004Princeton,Slanina2014Oxford}.

The long range dependence, considered as slowly decaying auto-correlation for  absolute returns, is a characteristic property of the empirical financial asset return series. The debate whether this slow decay corresponds to long range dependence is still ongoing    \cite{Giraitis2007,Giraitis2009,Conrad2010,Arouri2012,Giraitis2012,Tayefi2012}. Econometricians tend to the conclusion that the statistical analyses cannot be expected to provide a definite answer concerning the presence or absence of long-range dependence in asset price returns \cite{Lo1991Econometrica,Willinger1999FinStoch,Mikosch2003}. A deeper understanding of microscopic and macroscopic market forces is needed to build models of the financial markets reproducing the main stylized facts including the long-range dependence. Using a general agent-based stochastic model
\cite{Gontis2014PlosOne} various statistical properties of the financial markets, including high volatility return intervals
\cite{Yamasaki2005PNAS,Wang2006PhysRevE,Wang2008PhysRevE,Bunde2011EPL,Bunde2014PRE}, have been reproduced \cite{Gontis2015arXiv}.

Behavioral finance is frequently seen as an alternative view to the financial market efficiency as it relates the large price fluctuations to the animal spirits, for example, human brain bugs or herding tendencies, see recent books \cite{Shiller2015IE,Scheinkman2014CUP}. Human conformity, from the statistical point of view is equivalent to the herding, which we consider as statistically dominant in the endogenous dynamics of the financial markets. This endogenous dynamics  in interaction with the exogenous noise reproduces the main most general statistical properties of the real markets \cite{Gontis2015arXiv}. Namely, we follow a basic idea from the statistical physics that the individual intricacies of each trader are not so statistically important. The traders can be assumed boundedly rational, as their rationality is just too heterogeneous to be considered as statistically meaningful for the macroscopic outcome of the financial market. Thus we consider the global herding interaction of agents, quantified by A. Kirman’s transition rates \cite{Kirman1993QJE} in one step Markov chain, as an essential ingredient in the consentaneous agent based and stochastic modeling \cite{Gontis2014PlosOne}. This leads us to the financial market model with bursting endogenous fluctuations, which statistically matches the empirical data of various markets, various assets and can produce the large price movements on a longer time scales than it is allowed by the financial market efficiency hypothesis \cite{Slanina2014Oxford}.

Here we concentrate on the analysis of various noises included into the proposed modeling in order to reveal their statistically meaningful impact. Short overview of Markov processes with the power-law behavior and the long-range dependence is given in Section 2. In Section 3, we present a short description of the agent based model, which could be considered as some generalization of the models proposed by other authors. Description and numerical results, showing how the endogenous and exogenous fluctuations interplay, are presented in Section 4. Section 5 is devoted to the modeling and analyzes of the volatility return intervals seeking to reveal the contribution of various noises and, Section 6 resumes with the concluding remarks.  

\section{Markov processes with power-law behavior and long range dependence}
\begin{figure}[ht]
\centering
\includegraphics[width=0.45\textwidth]{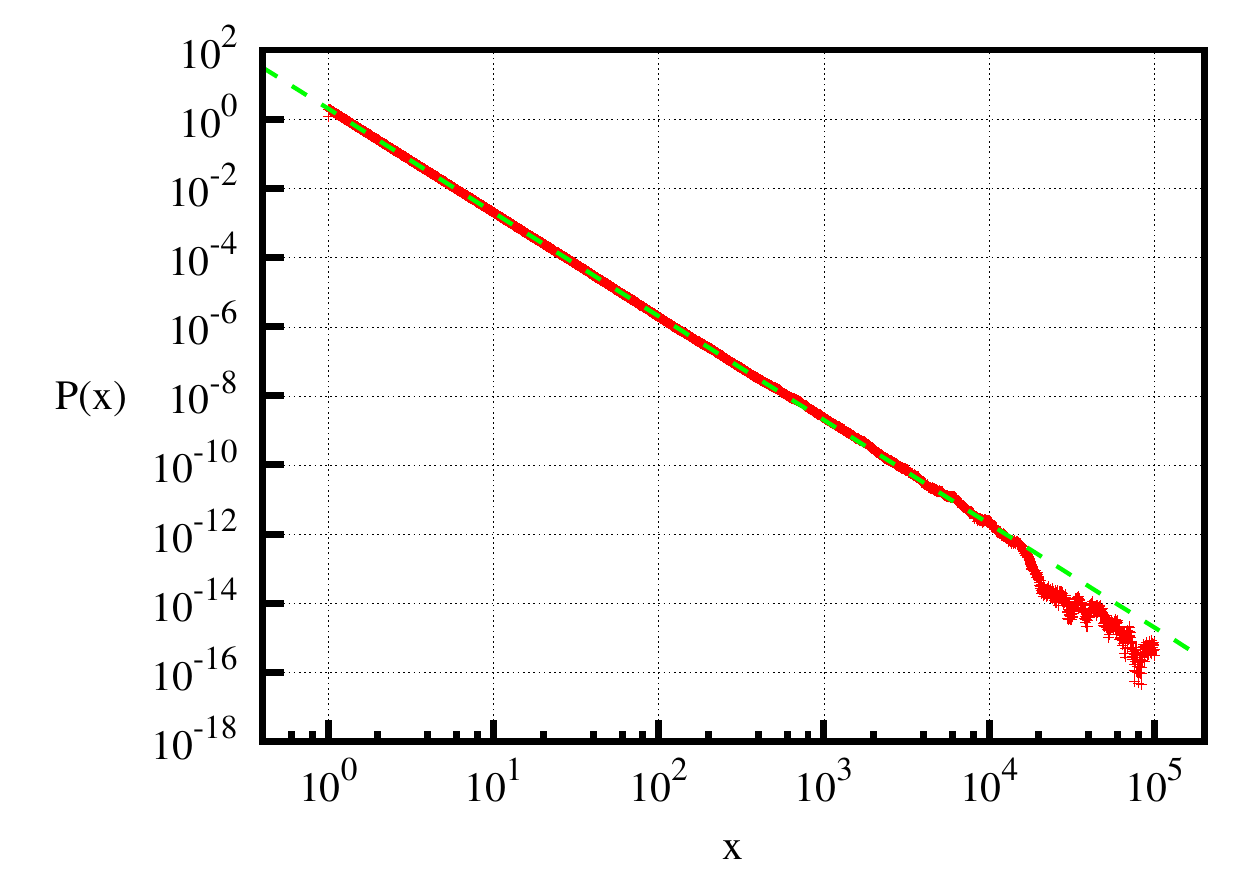}
\includegraphics[width=0.45\textwidth]{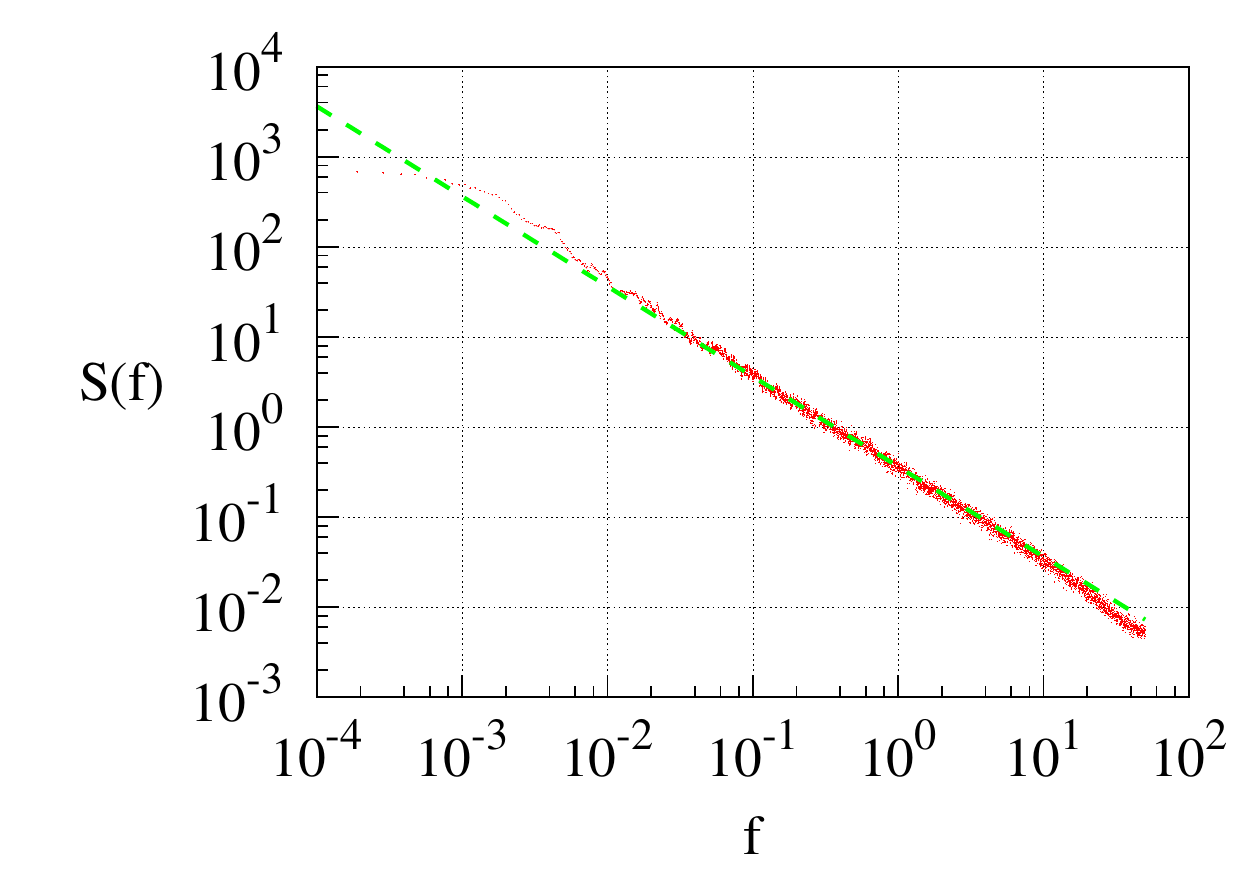}
\caption{\label{fig1} Numerically calculated (solid lines) stationary PDF (a) of the signal $\vert x \vert$ , Eq. (\ref{eq:SDE1}),  and PSD (b). Power-law fit is given as dashed straight lines.}
\end{figure}

Historically  the fractional Brownian motion has become the prevailing mathematical construct in understanding of the self-similarity and the long range dependence observed in the financial markets \cite{Baillie1996JE}. Here we argue once again that the nonlinear stochastic differential equations (SDE) can serve as an alternative mechanism dealing with the property of the long range dependence. Let us start from the most simple case of SDE exhibiting one over $f$ noise
\begin{equation}
\mathrm{d} x=(1+x^2)^{3/4} \mathrm{d} W.
\label{eq:SDE1}
\end{equation}
The signal $x$ defined by the SDE (\ref{eq:SDE1}) exhibits $q$-Gaussian stationary power-law probability density function (PDF) with the power-law tail $P(x)\sim x^{-3}$ and the power spectral density (PSD) $S(f)\sim \frac{1}{f}$, see \cite{Ruseckas2011PRE} and the numerical results given in Fig. \ref{fig1}.
   
This special case of Markov process with the long range dependence can be generalized by the class of nonlinear SDE's given as \cite{Ruseckas2014JStatMech},
\begin{equation}
\mathrm{d}x=(\eta-\frac{\lambda}{2})
 x^{2\eta-1}\mathrm{d} t +x^{\eta} \mathrm{d} W,
\label{eq:SDE2}
\end{equation} 
having just two parameters:  $\eta$ as exponent of noise multiplicativity and $\lambda$ as exponent of power-law PDF. It was justified by various methods that this class of SDE generates the time series with the power-law behavior of PDF and PSD \cite{Ruseckas2014JStatMech},
\begin{equation}
P(x)\sim x^{-\lambda},\quad S(f)\sim \frac{1}{f^\beta},\quad \beta=1+\frac{\lambda-3}{2\eta-2}.
\end{equation}
The range of frequencies $f$, where this type of PSD takes place, is defined by the limits $x_{\mathrm{min}}$ and $x_{\mathrm{max}}$ of $x$ diffusion restriction in Eq. (\ref{eq:SDE2}), \cite{Ruseckas2014JStatMech}, 

\begin{eqnarray}
& x_{\mathrm{min}}^{2 \eta -2}  << 2\pi f<< x_{\mathrm{max}}^{2 \eta -2},\quad for \quad \eta>1, \\
& x_{\mathrm{max}}^{2-2 \eta} \ll 2\pi f\ll x_{\mathrm{min}}^{2-2 \eta },\quad for \quad\eta<1.
\end{eqnarray}
Note that the class of Markov processes Eq. (\ref{eq:SDE2}) is not a long-range dependent in strict definition as frequencies are limited from the side of low values and time lags of auto-correlation from the side of high values. Nevertheless, having in mind that in real processes such limits are always natural it is rational to assume that this class of SDE's can be considered as possible description of the so called long-range  dependence in the financial markets. 

\begin{figure}[ht]
\centering
\includegraphics[width=0.65\textwidth]{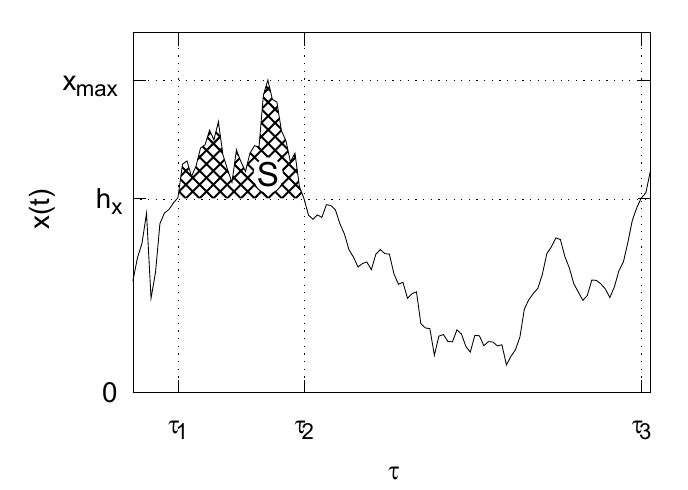}
\caption{\label{fig2} An example of signal $x(t)$, generated by Eq. (\ref{eq:SDE2}).}
\label{fig:burst-explanation-of-variables}
\end{figure}

The class of stochastic processes, Eq. (\ref{eq:SDE2}) has much more power-law properties. First of all we are interested in bursting behavior of the signal, having direct relation to the risk assessment in the financial markets. Let us define a burst as a part of signal, $x(t)$, lying above the fixed threshold, $h_x$, see Fig. \ref{fig2} as a graphical illustration. The time interval, $\tau=\tau_2-\tau_1$, here  
serves as a definition of the burst duration and the burst return interval can be defined as $T=\tau_3-\tau_1$. One can consider inter-burst time as, $\theta=\tau_3-\tau_2$.

It is possible to use the first hitting (passage) time framework \cite{Borodin2002Birkhauser,Jeanblanc2009Springer,Gardiner2009Springer,Redner2001Cambridge} to derive PDF of burst duration, $\tau$, see \cite{Gontis2012ACS}. The asymptotic behavior of $\tau$ PDF can be written in rather transparent form
\begin{eqnarray}
p_{h_x}^{(\nu)}(\tau) &\sim \tau^{-3/2} , \quad when\quad 0 < \tau \ll \frac{2}{(\eta-1)^2 h_x^{2(\eta-1)}j_{\nu,1}^2},\\
p_{h_x}^{(\nu)}(\tau) &\sim \frac{1}{\tau}\exp\left(-\frac{(\eta-1)^2 h_x^{2(\eta-1)}j_{\nu,1}^2 \tau}{2}\right) , \quad when\quad  \tau \gg \frac{2}{(\eta-1)^2 h_x^{2(\eta-1)}j_{\nu,1}^2}.
\label{eq:burst}
\end{eqnarray}
Here, $\nu=\frac{\lambda-2\nu+1}{2(\eta-1)}$, and $j_{\nu,1}$ is a first zero of a Bessel function of the first kind. The power-law behavior with exponent $3/2$ in Eq. (\ref{eq:burst}) is consistent with the general theory of the first-passage times in one-dimensional stochastic processes \cite{Redner2001Cambridge,Jeanblanc2009Springer}. It is possible to show by the numerical calculations that exponent $3/2$ is retained in the PDFs of burst return interval, $T$, and of inter-burst time, $\theta$, see \cite{Kaulakys2009ICNF,Kaulakys2009JStatMech}. We will use these power-law properties of the signal, generated by Eq. (\ref{eq:SDE2}), to argue the use of the nonlinear SDE in the modeling of the financial markets.

\section{Generalized agent based herding model of the financial markets}
We aim to construct a realistic model of the financial market based on the power-law properties of nonlinear SDE, bridging it with microscopic behavior of the financial agents. One has to recognize that this task is naturally very ambitious and there are many other attempts to build the agent based models of the financial markets \cite{Kirman2001SNDE, Pakkanen2010MathFinanEcon,Chakraborti2011RQUF2,Chen2012IRFA,Feng2012PNAS,Krause2012PhysRevE,
Frederick2013PNAS,Ortisi2013,Abergel2013NEW}. The variety of the already proposed agent based models confirms the ambiguity of such an approach. From our point of view the analytical treatment of the agent based models and detailed comparison of results with the empirical data  should be employed making the selection between various alternatives. Thus we start from the most simple version of agent interactions and add some new features when better adjustment to the empirical data is needed. 

As a first step we selected Kirman's herding interactions of two widely accepted groups of agents: chartists (speculative traders) and fundamentalists. Fundamentalists buy stocks when market price is lower than fundamental value and sell when market price is higher. Chartists are high frequency speculative traders making heterogeneous decisions regarding the further market price movement. It was shown that such agent system is able to catch up the long range dependence of volatility \cite{Kirman2001SNDE}.  The consideration of the same agent system was proposed in \cite{Alfarano2005CompEco}, which we adopted in a more appropriate form for our purposes and derived  macroscopic (stochastic) equation for the ratio of chartists and fundamentalists $x$, see \cite{Kononovicius2012PhysA}. The main innovation was to introduce the variable trading activity of the agent system as some feedback from macroscopic state.
This increases the exponent of multiplicativity $\eta$ in macroscopic SDE for the ratio $x$. The SDE for $x$ in the region of high values can be approximated by the class of SDE (\ref{eq:SDE2}) with parameters: 
\begin{eqnarray}
& \eta=\frac{3+\alpha}{2}, \label{eq:eta}\\
& \lambda=\varepsilon_2+\alpha+1. \label{eq:lambda}
\end{eqnarray}
Here $\alpha$ is a feedback parameter and $\varepsilon_2$ is idiosyncratic transition rate of chartists to fundamental behavior divided  by herding parameter $h$, see \cite{Kononovicius2012PhysA} for details. As in this simplified model $x$ has a meaning of the long-term absolute return, its' power law behavior is informative about validity of such approach. Note that exponent of PSD, $\beta$, can be written as
\begin{equation}
\beta=1+\frac{\varepsilon_2+\alpha-2}{1+\alpha}.
\label{eq:beta}
\end{equation}
Having in mind that the exponent of empirical absolute return power-law PDF is approximately 4, the corresponding parameter in Eq. (\ref{eq:SDE2}) has to be $\lambda \simeq 4$. The empirical analyzes suggests that the trading activity is
proportional to the square of the absolute returns, this implies $\alpha=2$ and from Eq. (\ref{eq:lambda}) the value of $\varepsilon_2 \simeq 1$ follows. This from Eq. (\ref{eq:beta}) implies a value of $\beta$ in the region $1 \lesssim \beta \lesssim 1.5$, which complies with the long-range dependence, but is too high in comparison with empirical value below $1$. It was shown in \cite{Kononovicius2012PhysA} that this model is compatible with multifractal behavior of absolute return but it is not possible to reproduce empirical PDF and PSD of absolute return with the same set of parameters.

There is one more problem making comparison with the empirical long-range dependence of absolute return related with two scales of long-range dependence, as empirical PSD has at least two values of $\beta$, $\beta_1\backsimeq0.7$ and $\beta_2\backsimeq0.3$. To solve this problem we have to introduce one more source of the return fluctuations acting in shorter time scales than chartist fundamentalist herding dynamics. Thus we modify this herding model by dividing  chartists into two groups: optimists and pessimists.

Now we have slightly more sophisticated herding model with three groups of agents: fundamental ($f$), optimistic ($o$) and pessimistic ($p$), where relative population of each group, $n_i$, varies under constraints $\sum_{i} n_i=1$. As usually \cite{Lux1999Nature,Alfarano2005CompEco,Samanidou2007RepProgPhys,Cincotti2008CompEco,
Feng2012PNAS}, the fundamental traders assume that the price should be based on the market fundamentals quantified by some value $P_f$.  The previous chartist ($c$) trading strategy now consists of optimistic and pessimistic trading, where optimists always buy and pessimists always sell. Let us define the  excess demands, $D_i$, for both  fundamental and  chartist strategies, as was proposed in \cite{Alfarano2005CompEco},
\begin{equation}
D_f = n_f \left[ \ln P_f - \ln P(t) \right] , \quad D_c = r_0 (n_o - n_p) = r_0 n_c \xi ,
\end{equation}
where $P(t)$ is the current market price of asset, $r_0$ describes relative impact of chartists and $\xi = \frac{n_o - n_p}{n_c}$ is their average mood. The balance of demands $D_f$  and $D_c$  defines the log-price \cite{Alfarano2005CompEco,Gontis2014PlosOne}:
\begin{equation}
p(t) = \ln \frac{P(t)}{P_f} = r_0 \frac{n_c}{n_f} \xi = r_0 \frac{1-n_f}{n_f} \xi . \label{eq:pdefin}
\end{equation}

As the empirical PSD of the absolute returns demonstrates, it is rational to assume optimists and pessimists as being high frequency traders, trading between themselves on the intraday time scales $H$ times more frequently than trading with fundamentalists. This allows us to simplify the agent population $n_i$, dynamics  arising from one step transition $i \rightarrow j$ rates proposed by Kirman \cite{Kirman1993QJE}: 
\begin{equation}
\mu_{ij} = \sigma_{ij} + h n_j N , \label{eq:nonext}
\end{equation}
where $\sigma_{ij}$ describes idiosyncratic switching tendency, while $h$ term quantifies influence of peers, $n_j N$. Note that number of peers in the global pairwise coupling of agents is proportional to the total number of agents $N$. 

The following symmetric relationships ($\sigma_{op}=\sigma_{po}=\sigma_{cc}$, $h_{op}=H h_{fc}=H h$), ($\sigma_{pf}=\sigma_{of}=\sigma_{cf}$), ($\sigma_{fp}=\sigma_{fo}=\sigma_{fc}/2$ and $h_{fp}=h_{fo}=h$) based on general understanding of the financial market dynamics greatly simplify the model. The assumption that fundamentalists are the long-term traders whereas the chartists are the short-term traders means that ($H \gg 1$, $\sigma_{cc} \gg \sigma_{cf}$ and $\sigma_{cc} \gg \sigma_{fc}$). Under these assumptions the dynamics are well approximated by two almost independent SDEs \cite{Kononovicius2013EPL,Gontis2014PlosOne}, resembling the original SDE from the two state herding model \cite{Kirman1993QJE,Alfarano2005CompEco}:
\begin{eqnarray}
& \rmd n_f = \frac{(1-n_f) \varepsilon_{cf} - n_f \varepsilon_{fc}}{\tau(n_f)} \rmd t + \sqrt{\frac{2 n_f (1-n_f)}{\tau(n_f)}} \rmd W_{f} , \label{eq:nftau}\\
& \rmd \xi = - \frac{2 H \varepsilon_{cc} \xi}{\tau(n_f)} \rmd t + \sqrt{\frac{2 H (1-\xi^2)}{\tau(n_f)}} \rmd W_{\xi}. \label{eq:xitau}
\end{eqnarray}
Here $\tau(n_f)$, as the above mentioned  macroscopic feedback, is the inter-event time, whereas  $W_{f}$ and  $W_{\xi}$ 
  are independent Wiener processes. Note that in the above equations we  scale model parameters, $\varepsilon_{cf} = \sigma_{cf} / h$, $\varepsilon_{fc} = \sigma_{fc} / h$ and $\varepsilon_{cc} = \sigma_{cc} / (H h)$, as well as time $t_s = h t$, omitting subscript $s$ in the equations.

Inter-event time $\tau(n_f)$ takes the following form
\begin{equation}
\frac{1}{\tau(n_f)}= \left( 1 + a_{\tau} \left| \frac{1-n_f}{n_f} \right| \right)^{\alpha}. \label{eq:taunfxi}
\end{equation}
This form is inspired by the empirical analysis  \cite{Gabaix2003Nature,Farmer2004QF,Gabaix2006QJE,Rak2013APP}, where  the trading activity is proportional to the square of absolute return (thus $\alpha=2$). Note that the given form depends on the long-term component of return in the proposed model, $x=\frac{1-n_f}{n_f}$. 

Equations (\ref{eq:nftau}-\ref{eq:taunfxi}) form the complete set for macroscopic description of the endogenous agent dynamics and together with Eq.  (\ref{eq:pdefin}) can be considered as a model of the financial markets. Eq. (\ref{eq:nftau}) written for the new variable $x=\frac{1-n_f}{n_f}$ in the region of high values of variable belongs to the class of the non-linear SDEs,  reproducing power-law statistics: PDF and PSD \cite{Kaulakys2005PhysRevE,Ruseckas2014JSM}. It was shown in \cite{Kononovicius2013EPL} that such a three state herding model is able to reproduce the long-range dependence of absolute return with two exponents of PSD. Nevertheless, the values of these exponents are too high and there is no opportunity to adjust the endogenous model of the financial markets to the empirical data. We have to conclude that something very essential is missing in this modeling. 

\section{Interplay between endogenous and exogenous fluctuations} 

In the previous modeling of the financial markets by the nonlinear SDE \cite{Gontis2006JStatMech,Gontis2007PhysA,Gontis2008PhysA,Gontis2010PhysA} it was shown that the additional stochastic process driven by the class of SDE's  can solve the problem of too high exponents in PSD. Thus we consider the opportunity to introduce one more source of the return fluctuations related with the exogenous information or order flow noise. In other words, we do doubt in modeling of the financial markets based just on the endogenous dynamics of agents, thus combine it with some phenomenological description of high frequency exogenous noise.

It is widely accepted to describe the dynamics of asset price $S(t)$ in the financial markets by standard model \cite{Jeanblanc2009Springer} written as geometric Brownian motion
\begin{equation}
\rmd S_t = S_t(\mu_t \rmd t + \sigma_t \rmd W),
\label{eq:Standard}
\end{equation}
where $\mu_t$ stands for the slowly varying trend, $\sigma_t$ for the slowly varying volatility and $W$ for the Brownian motion (Wiener process). The physical interpretation of the Wiener process in Eq. (\ref{eq:Standard}) was recently considered as a motion of the financial Brownian particle colliding with the flow of limit orders in the real financial market \cite{Takayasu2015PRE}. This is one more argument for us to interpret $W$ as a source of high frequency exogenous noise incorporating fluctuations of information flow through the order flow. It is natural to assume that $\sigma_t$ reflects much slower varying endogenous state of agent system we discussed in the previous section. The long-range trend in Eq. (\ref{eq:Standard}) is described by slowly varying $\mu_t$, which could be considered as reflecting slow movement of the fundamental price or persistent trend of the exogenous noise.
Seeking to simplify our task it is rational to restrict the study with modeling of return leaving out the long-term movement of price $S$, which is too much related to the  exogenous information quantified by $\mu(t)-\frac{1}{2} \sigma_t^2$. Thus we define from Eq. (\ref{eq:Standard}) the short term return $r_{\delta}(t)$ in a very short time period $\delta$, where $\sigma_t$ can be assumed as constant,
\begin{equation}
r_{\delta}(t) = \sigma_t \omega_t.
\label{eq:return}
\end{equation}
Here $\omega_t$ is a Gaussian noise with zero mean and unit variance. Note that Eq. (\ref{eq:return}) coincides with return definition in the family of autoregressive conditional heteroskedasticity (ARCH)  models.
The volatility $\sigma_t$ is assumed in \cite{Gontis2014PlosOne} as a linear function of the
absolute endogenous log price $\vert p(t) \vert$ defined in Eq. (\ref{eq:pdefin}) 
\begin{equation}
\sigma_t = b_0(1+ a_0 \vert p(t) \vert),
\label{eq:defvolatil}
\end{equation}
here $b_0$ serves as a normalization
parameter, when we normalize model and empirical series in the same way, $b_0=1$; $a_0$ determines the impact of endogenous dynamics on
the observed time series. The model, defined by Eqs.~(\ref{eq:return})
and (\ref{eq:defvolatil}), comprises both the endogenous dynamic part described by
$\sigma_t$ and the exogenous noise part described by $\omega_t$.

Substitution of endogenous price $p(t)$, Eq. (\ref{eq:pdefin}), calculated by Eqs. (\ref{eq:nftau} - \ref{eq:taunfxi}) for $n_f$ and $\xi$, into Eqs. (\ref{eq:return} - \ref{eq:defvolatil}) completes the model including the endogenous and exogenous fluctuations. It is possible to demonstrate \cite{Kononovicius2015PhysA} that such model resembles the versions of non-linear GARCH(1,1) models \cite{Engle1986ER,Higgins1992IntEcoRev}. At the same time the proposed model is based on the multifractal point process \cite{Gontis2006JStatMech,Gontis2007PhysA} and should be considered as an alternative to the modeling by Hawkes self-excited point process model \cite{Filimonov2015QF}.

Phenomenological and pure stochastic models do not have sufficient insight into market dynamics.  
The advantage of the agent-based models is their insight into real life and real human behavior, which can be quantified by parameters. Such models can be modified, first, introducing the relationships between exogenous and endogenous fluctuations and, second, adjusting their time scales.

The selected time window $\delta$ is limited by the requirement that the change of $\sigma_t$ has to be inconsiderable. Nevertheless, one can calculate the return in longer time window $\Delta$ by summing up the consecutive short-term returns $r_{\delta}(t)$
\begin{equation}
r_{\Delta}(t)=\sum_{i=1}^{\Delta/\delta}r_{\delta}(t+i \delta).
\label{eq:returnDelta}
\end{equation}
Thus the resultant series of return $r_{\Delta}(t)$ already incorporate endogenous and exogenous fluctuations of the financial markets. 

The interplay of endogenous and exogenous fluctuations might be influenced by the intro-day fluctuations observed in real markets. To account for the daily pattern a time dependence was introduced \cite{Gontis2014PlosOne}
 into parameter $b_0$, i.e., 
\begin{equation}
b_0(t)=b_0 \exp [-(\{t \mathrm{mod} 1 \} -0.5)^2/w^2]+0.5,
\label{eq:b0}
\label{eq:intraday}
\end{equation}
  where $w$ quantifies the width of intra-day fluctuations. 
Though the model  has been designed to reproduce the long-range dependence and power-law behavior of absolute return, from our point of view, it suites very well to investigate the interplay of endogenous and exogenous noise reflected in the empirical data of absolute return PDF, PSD and return intervals \cite{Yamasaki2005PNAS,Wang2006PhysRevE,Wang2008PhysRevE}.

In \cite{Gontis2015arXiv} we already adjusted parameters of given model to the empirical data of NYSE and FOREX. Good agreement was achieved for the set of parameters: $\delta=1/390$ trading day $=3.69$ min. which is equivalent to 1 NYSE
trading minute, $\varepsilon_{cf}=1.1$ and $\varepsilon_{fc}=3$, $\varepsilon_{cc}=3$, $H=1000$ 
adjusts the PSDs of the empirical and model time series, $a_0=1$ and
$a_{\tau}=0.7$ are the empirical parameters defining the sensitivity
of market returns and trading activity to the populations of agent
states, $\alpha=2$ is selected on the basis of empirical analysis
\cite{Gabaix2003Nature,Farmer2004QF,Gabaix2006QJE,Rak2013APP}, and $h=0.3\times10^{-8}
s^{-1}$ is the main time-scale parameter that adjusts the model
to fit the real time-scale.
\begin{figure}[ht]
\centering
\includegraphics[width=0.95\textwidth]{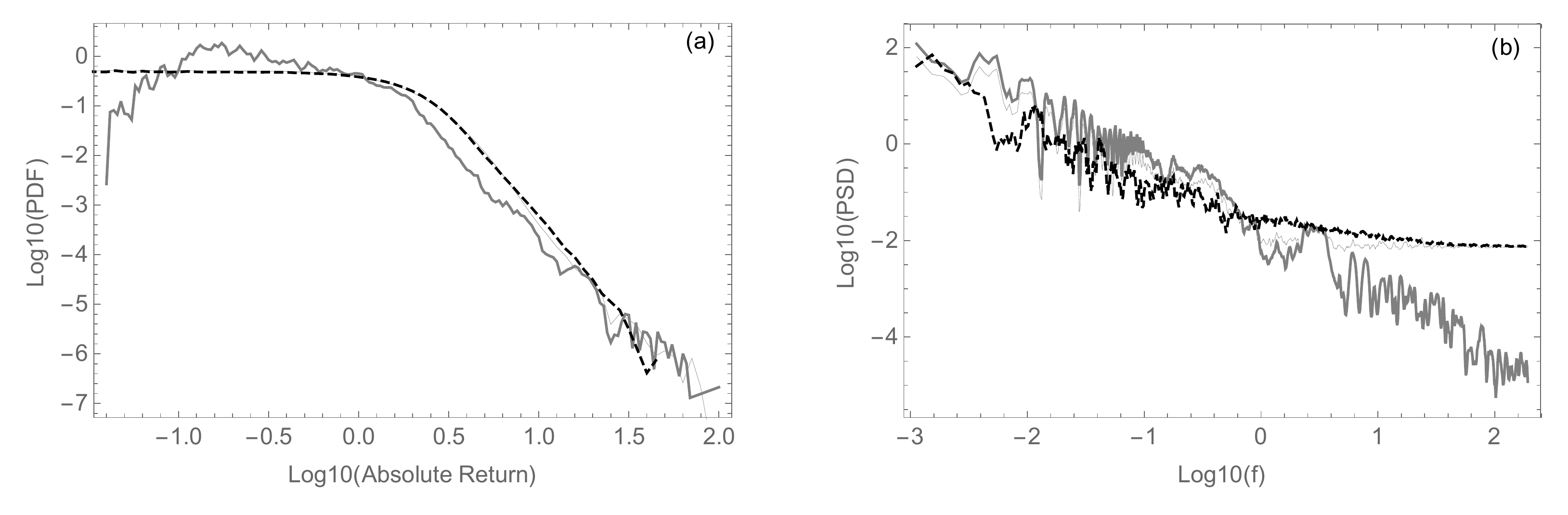}
\caption{\label{fig3} Numerical results exhibiting different components of the proposed model of one minute ($\Delta=\delta=1/390$ trading day) return. (a) stationary PDF of absolute return: just of $n_c/n_f=(1-n_f)/n_f$ in time steps $\delta$ (solid gray line); of $\mid r_\delta(t) \mid$, with constant $\xi\equiv 1$ and $b_0\equiv 1$ (thin gray line); and with constant $b_0\equiv 1$ only (dashed black line); (b) PSD for the same variables as in (a). All other model parameters are the same in this contribution.}
\end{figure}

It was shown in \cite{Gontis2015arXiv} that the proposed model of the financial markets with the same set of parameters reproduces PDF and PSD of absolute return and statistics of volatility return intervals for all assets analyzed from the NYSE and FOREX markets with return definition times $\Delta$ ranging from one minute to one month. Here we demonstrate how these statistical properties depend on the constituent noises of proposed model. First, the long-term chartist fundamentalist dynamics is described by  ratio $n_c/n_f=(1-n_f)/n_f$ defined by Eq. (\ref{eq:nftau}). Second, keeping $\xi$ and $b_0$ constants we investigate the interaction of the long-term endogenous dynamics with exogenous noise by analyzing $\mid r_\delta(t) \mid$ and $\mid r_\Delta(t) \mid$. Third, we analyze the absolute return series, when optimist-pessimists dynamics $\xi(t)$ is switched on as well. 

In Fig. \ref{fig3} we demonstrate PDF (a) and PSD (b) of these three time series, when $\Delta=\delta=1/390$ of trading day. One can observe that the power-law behavior of these first and second order statistics first of all is defined by $n_c/n_f$ dynamics represented by solid gray lines in Fig. \ref{fig3}. When we add the exogenous noise (thin gray line), it slightly transforms PDF, making it smoother, and transforms PSD in high frequency area making it flat as white noise. When we add the high frequency fluctuations of $\xi$ (dashed black line), they are not reflected in PDF, which practically coincides with the previous case, but impact PSD, which demonstrates smooth transition from low frequencies to high frequencies.
\begin{figure}[ht]
\centering
\includegraphics[width=0.95\textwidth]{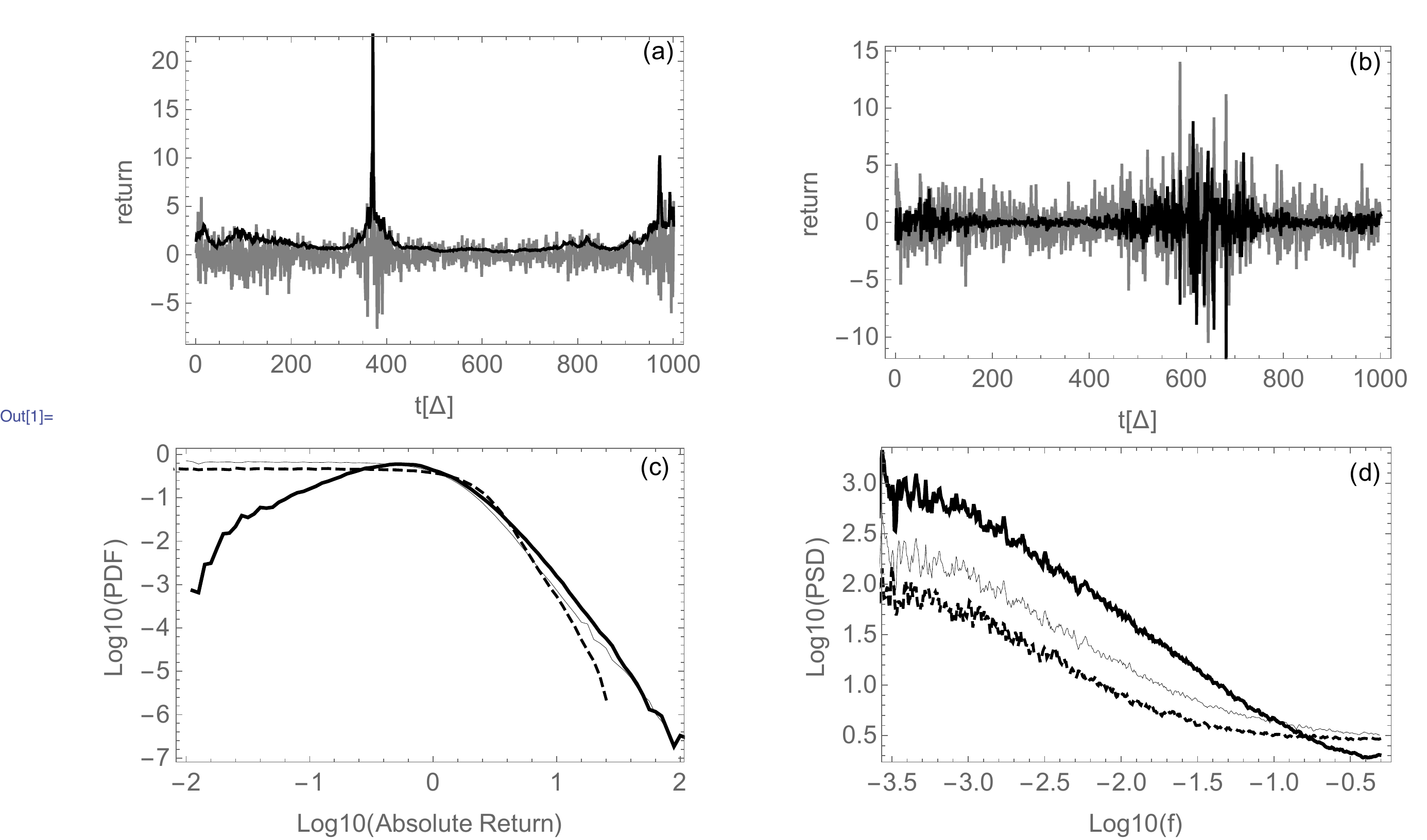}
\caption{\label{fig4} Numerical results exhibiting different components of the proposed model of daily return with $b_0\equiv 1$. (a) example of $n_c/n_f=(1-n_f)/n_f$ time series (black line) in comparison with return $r_{\Delta}(t)$ (gray line) calculated keeping $\xi$ constant; (b) example of $n_c/n_f \xi(t)$ time series (black line) in comparison with return $r_{\Delta}(t)$ (gray line) calculated with chartist's mood $\xi$;  (c) stationary PDF of absolute return: just of $n_c/n_f=(1-n_f)/n_f$ (solid black line), of $\mid r_\Delta(t) \mid$, with constant $\xi\equiv 1$ (thin line), and $\mid r_\Delta(t) \mid$ with constant $b_0\equiv 1$ only (dashed black line); (d) PSD for the same variables as in (c), $n_c/n_f$ upper, with $\xi$ lower. All other model parameters are the same in this contribution.}
\end{figure}

In Fig. \ref{fig4} we demonstrate behavior of the same model components as in Fig. \ref{fig3}, but with daily definition of return, $\Delta=1$  day. Extracts of return time series are given in sub-figures (a) and (b), where  modulating signals (black lines) are compared with returns accounting exogenous noise (gray lines): in sub-figure (a) $\xi\equiv 1$ and in sub-figure (b) dynamics of chartists is included. Stationary PDFs of return are given in sub-figure (c), where solid black line illustrates $n_c/n_f$, thin gray line adds exogenous noise and dashed line includes $\xi$ fluctuations as well. Note that high frequency endogenous fluctuations are able to decrease the exponent of power-law PDF, when time scale $\Delta$ exceeds characteristic time of these fluctuations.  We plot PSD of absolute returns in sub-figure (d), which demonstrates strong dependence of power-law exponents on both high frequency and exogenous fluctuations. The same scaling of empirical and model PSD of returns with $\Delta$, see \cite{Gontis2015arXiv}, confirms that it is essential to account high frequency endogenous and exogenous fluctuations in the modeling of the financial markets as these fluctuations contribute to the long-term  statistical properties (PSD for low frequencies).  

\section{Statistical properties of volatility return intervals}

\begin{figure}[ht]
\centering
\includegraphics[width=0.65\textwidth]{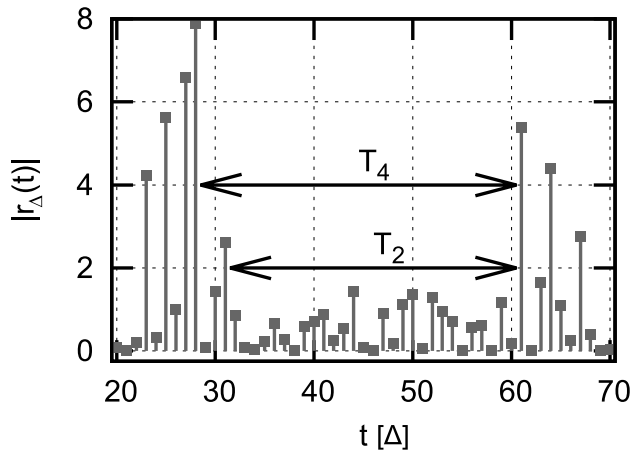}
\caption{\label{fig5} The definition of return intervals $T_q$. Return intervals $T_q$ between the
  volatilities of the price changes that are above a certain threshold
  q, measured in units of standard deviations of returns. Here two values of threshold $q=2$ and $q=4$ are shown in
  the time series of absolute return.}
\end{figure}

There is continuing extensive study of high volatility return intervals in the financial markets \cite{Yamasaki2005PNAS,Wang2006PhysRevE,Wang2008PhysRevE,Bunde2011EPL,Bunde2014PRE,Gontis2015arXiv,Denys2015arXiv}.
Let us discuss the statistical properties of return intervals from the perspective of the proposed model and various noises incorporated there. The power-law nature of Eqs. (\ref{eq:SDE2}) and (\ref{eq:nftau}) including burst duration power-law PDF with exponent $3/2$, from our point of view, is a key for understanding statistics of volatility return intervals. As in previous section we analyze the decomposition of noises incorporated into the proposed model seeking to reveal their input into the PDF of return intervals $T_q$. Recall definition of absolute return interval $T_q$, given in Fig. \ref{fig5}. Seeking to reveal scaling properties of $T_q$ PDF, we normalize series of $T_q$ by its' average $<T_q>$.

We investigate the discrete series of absolute return  $\mid r_{\Delta}(t) \mid$ calculated for four different   compositions of the model: a) $r_{\Delta}(t)=n_c/n_f$, Eq. (\ref{eq:nftau}); b) $r_{\Delta}(t)$ defined by Eqs. (\ref{eq:returnDelta}) and (\ref{eq:return}), $\sigma_t = 1+ a_0 \vert n_c/n_f \vert$; c) the same as previous with $\sigma_t = 1+ a_0 \vert n_c/n_f \xi_t \vert$ including $\xi$ fluctuations; d) the same as previous including seasonality $b_0$ defined in Eq. (\ref{eq:b0}) and $\sigma_t = b_0 (1+ a_0 \vert n_c/n_f \xi_t \vert)$. 

\begin{figure}[ht]
\centering
\includegraphics[width=0.95\textwidth]{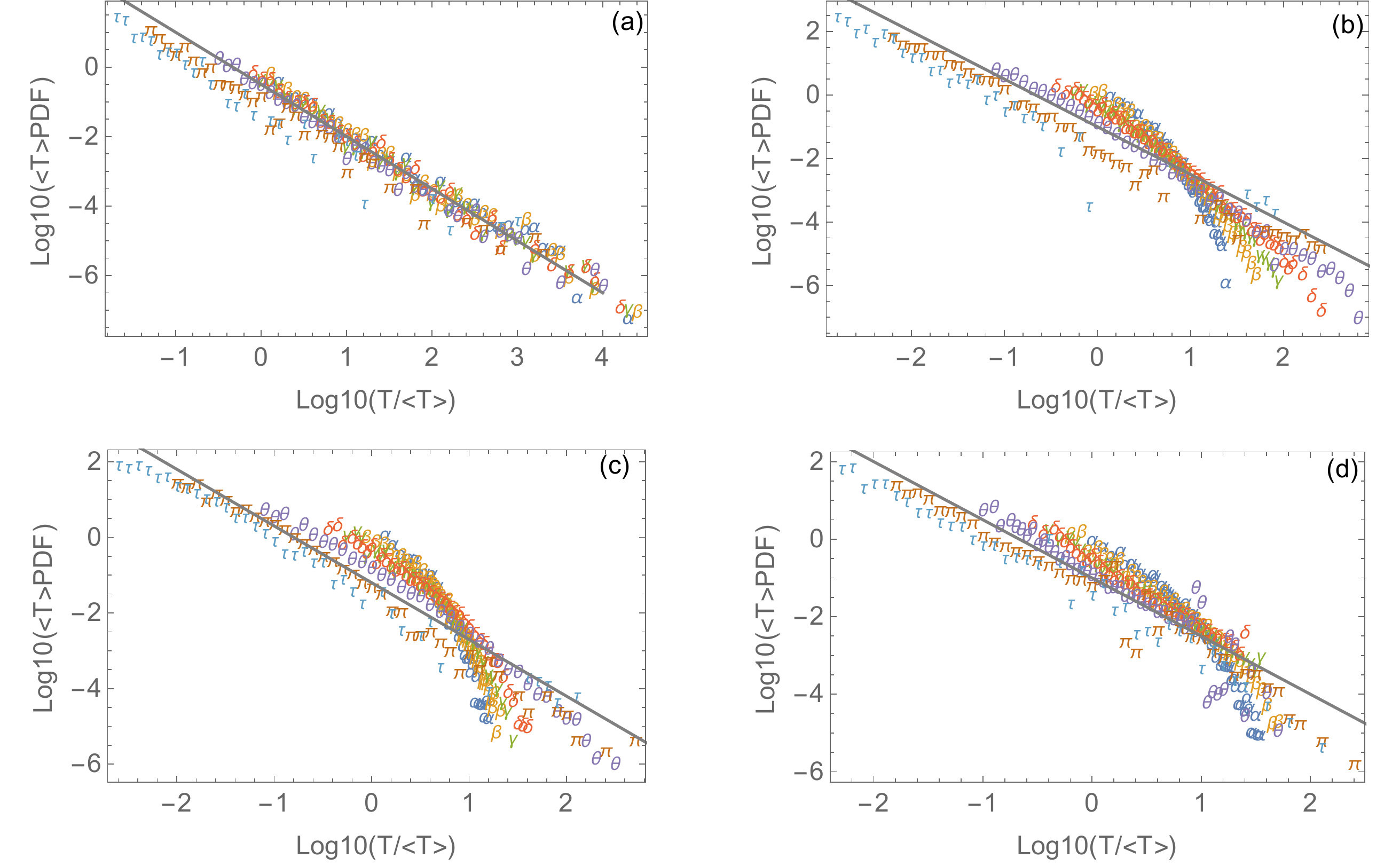}
\caption{\label{fig6} Scaled PDF of volatility return intervals $T_q$ for $\Delta=\delta=1/390$ trading day and normalized return thresholds $q=\{1.5, 2, 2.5, 3, 5, 10, 15\}$, where points of numerical histograms are denoted by Greek letters $\{\alpha, \beta, \gamma, \delta, \theta, \pi,
\tau\}$ for corresponding values of $q$. Distributions are plotted in log-log scale when $T_q$ values are normalized by series average $<T_q>$.  (a) PDFs just of $n_c/n_f=(1-n_f)/n_f$ calculated with time step $\Delta$;  (b) PDFs of $r_{\Delta}(t)$ defined by Eqs. (\ref{eq:returnDelta}) and (\ref{eq:return}), $\sigma_t = 1+ a_0 \vert n_c/n_f \vert$; (c) the same as previous PDFs with $\sigma_t = 1+ a_0 \vert n_c/n_f \xi_t \vert$ including $\xi$ fluctuations; d) the same as previous PDFs including seasonality $b_0$ defined in Eq. (\ref{eq:b0}) and $\sigma_t = b_0 (1+ a_0 \vert n_c/n_f \xi_t \vert)$. All other model parameters are the same in this contribution. The straight line guides the eye according 3/2 power-law. }
\end{figure}

In Fig. \ref{fig6} we plot 4 sub-figures of scaled $T_q$ PDFs for 4 mentioned compositions of the model return and  for $\Delta=\delta=1/390$ trading day. In each sub-figure PDFs for 7 values of threshold $q=\{1.5, 2, 2.5, 3, 5, 10, 15\}$ are given by corresponding Greek letters 
$\{\alpha, \beta, \gamma, \delta, \theta, \pi,
\tau\}$.  As one could expect, in the case (a) the sequence of returns is driven by one-dimensional stochastic process, which, according first hitting time theory \cite{Borodin2002Birkhauser,Jeanblanc2009Springer,Gardiner2009Springer,Redner2001Cambridge}, results in $3/2$ power-law of return interval PDF for all values of threshold $q$. Note that time step $\Delta=\delta=1/390$ for these return series is very small in comparison with $1/h$, therefore discreteness of the signal is not important. In sub-figure (b) the signal includes exogenous noise, which is responsible for the appearance of some exponential like  cutoff in PDF of return intervals. This effect is stronger for low thresholds $q$ and practically disappears for high values of $q$. When we add high frequency endogenous fluctuations $\xi$, (c) sub-figure, PDF of $T_q$ do not change considerably, as $\delta<<\frac{1}{h H}$. PDFs of $T_q$ do not experience considerable change even by dynamic seasonality, Eq. (\ref{eq:b0}), included in sub-figure (d). Therefore we conclude that interplay of exogenous noise with endogenous dynamics is a key factor resulting in observed power-law behavior cutoff of  $T_q$  PDF.  

\begin{figure}[h]
\centering
\includegraphics[width=0.95\textwidth]{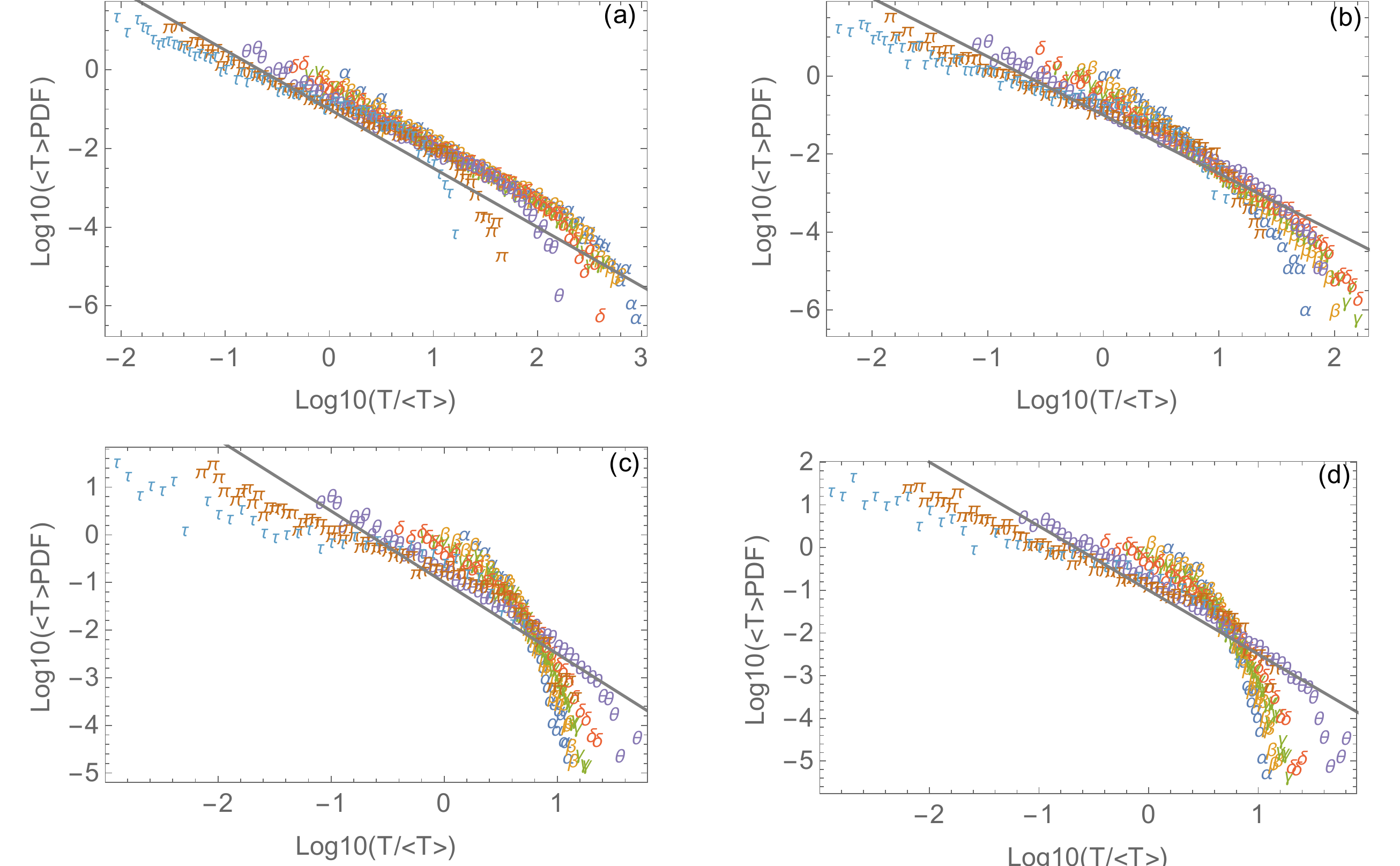}
\caption{\label{fig7} Scaled PDF of volatility return intervals $T_q$ for $\Delta=390 \delta=$ trading day and normalized return thresholds $q=\{1.5, 2, 2.5, 3, 5, 10, 15\}$, where points of numerical histograms are denoted by Greek letters $\{\alpha, \beta, \gamma, \delta, \theta, \pi,
\tau\}$ for corresponding values of $q$. Distributions are plotted in log-log scale when $T_q$ values are normalized by series average $<T_q>$.  (a) PDFs just of $n_c/n_f=(1-n_f)/n_f$ calculated by Eq. (\ref{eq:nftau}) with time step $\Delta$;  (b) PDFs of $r_{\Delta}(t)$ defined by Eqs. (\ref{eq:returnDelta}) and (\ref{eq:return}), $\sigma_t = 1+ a_0 \vert n_c/n_f \vert$; (c) the same as previous PDFs with $\sigma_t = 1+ a_0 \vert n_c/n_f \xi_t \vert$ including $\xi$ fluctuations; d) the same as previous PDFs including seasonality $b_0$ defined in Eq. (\ref{eq:b0}) and $\sigma_t = b_0 (1+ a_0 \vert n_c/n_f \xi_t \vert)$. All other model parameters are the same in this contribution. The straight line guides the eye according 3/2 power-law.}
\end{figure}

In Fig. \ref{fig7} we plot 4 sub-figures of scaled $T_q$ PDFs for the same 4 compositions of the model return with $\Delta=390 \delta=$ trading day. Values of thresholds and corresponding notations of PDFs are the same as in previous figure. In this case daily steps of continuous stochastic process $n_c/n_f$ result in discreetness of return series and even for simplest case, (a) without exogenous noise, deviations of $T_q$ PDFs from $3/2$ power-law appears for all values of threshold $q$ in the region of high $T_q$ values and for the highest thresholds in the region of low $T_q$ values. Nevertheless, for moderate threshold values some region of $T_q$ power-law $3/2$ behavior is still present. In sub-figure (b), where the signal includes exogenous noise, the deviations from $3/2$ power-law become more apparent and all PDFs scale nearly in the same functional form. When we add high frequency endogenous fluctuations $\xi$, (c) sub-figure, PDFs of $T_q$ deviate from $3/2$ power-law more considerably for high and low threshold values. For the $q=5$, $\theta$ points of $T_q$ PDF have clear part of $3/2$ power-law. This confirms the contribution of endogenous agent dynamics to the behavior of returns with daily definition, though the exogenous noise input seems more considerable in this case.  Dynamic seasonality included in sub-figure (d) does not change PDFs of $T_q$ considerably. Our numerical results based on the proposed model confirm the increasing impact of exogenous noise on statistics of volatility return intervals, when time window of return definition, $\Delta$, is wider.  

\begin{figure}[h]
\centering
\includegraphics[width=0.95\textwidth]{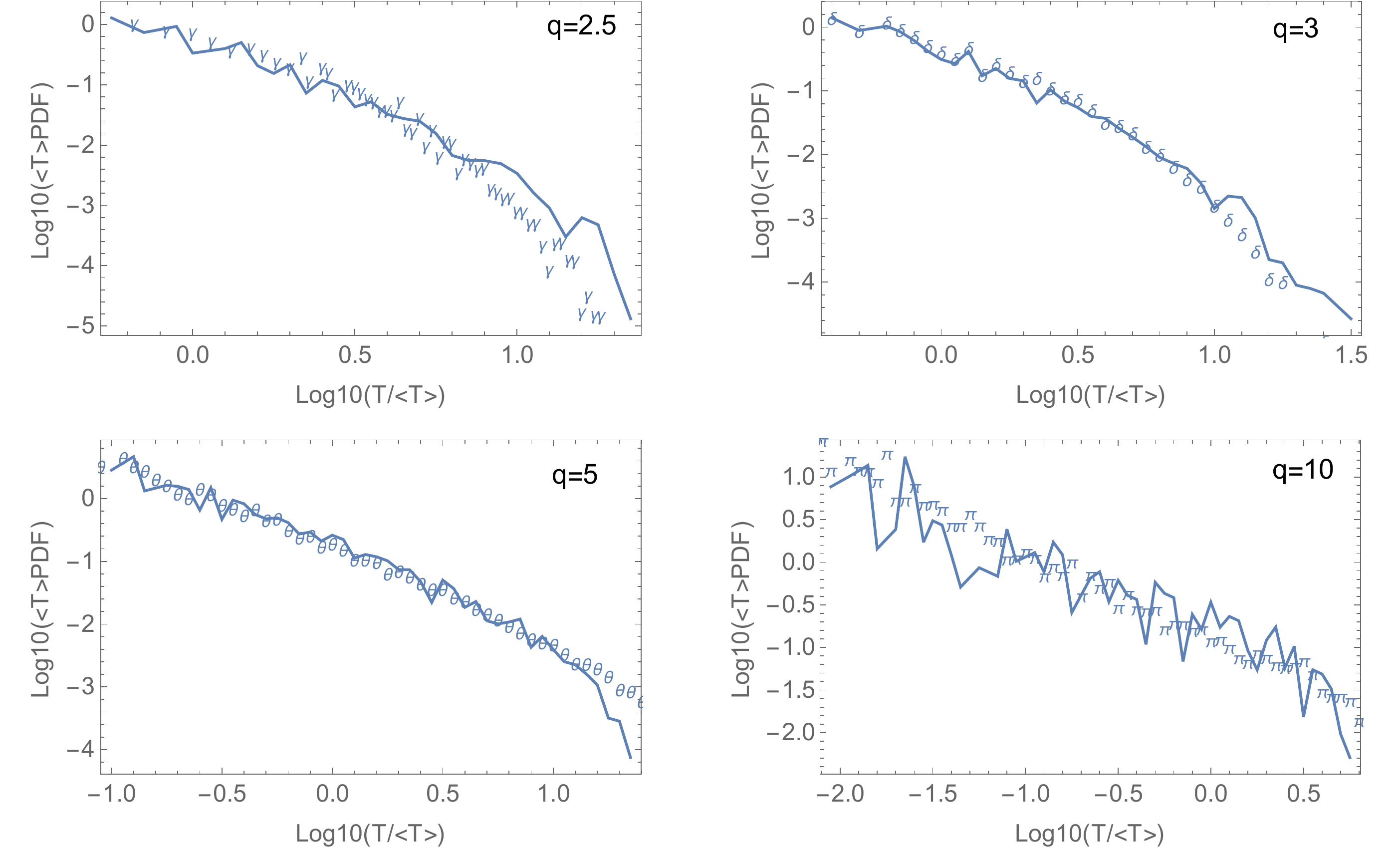}
\caption{\label{fig8} Comparison of scaled model and empirical PDFs of volatility return intervals $T_q$ for $\Delta=$ trading day. Points of numerical model histograms are denoted by Greek letters and empirical PDFs for 6 NYSE stocks calculated from normalized return
series are plotted by solid lines. Values of $q$ are given in the right corner of sub-figure. Distributions are plotted in log-log scale when $T_q$ values are normalized by series average $<T_q>$.   All other model parameters are the same in this contribution.}
\end{figure}

The detailed comparison of  statistical properties for return intervals generated by this this model and empirical data for NYSE stocks and FOREX exchange rates is given in \cite{Gontis2015arXiv}. In Fig. \ref{fig8} we provide only illustrative comparison of full scale model PDFs for daily volatility return intervals with PDFs calculated for absolute daily returns of 6 NYSE stocks. Series of return for 6 stocks obtained from Yahoo Finance have been normalized by standard deviation to the same PDF and have been joined to produce a single data set for absolute return intervals. Fig. \ref{fig8} plots comparison for 4 different threshold $q$ values given in sub-figures. Empirical and model PDFs coincide pretty well only some difference can be observed for $q=2.5$. The model does not reproduce empirical statistical properties of return intervals for low values of thresholds outside the power-law part of return PDF, where the complexity of behavior is very high. Probably there is a space for the model improvement taking into account the intraday fluctuations of trading activity more carefully. 

\section{Concluding remarks}
Herein, we analyzed how various noises impact statistical properties of absolute return in the financial markets. Earlier proposed model includes the long-term and high frequency endogenous fluctuations as well as the phenomenological exogenous high frequency fluctuations. This helps us to reveal numerically how these noises interplay contributing to various statistical properties including absolute return PDF, PSD and PDF of high volatility return intervals. 

First of all, we demonstrate that exogenous noise is essential in the modeling of long-range dependence of absolute return and volatility  as endogenous dynamics alone produces  too high values for the exponents of PSD.   

Second, we demonstrate that exogenous noise is very important for the understanding exponential like cutoff of high volatility return interval PDF, arising from the general hitting time theory of stochastic processes. 

And finally, our results confirm that the impact of exogenous noise increases with wider window $\Delta$ of return definition resulting in considerable deviations from $3/2$ power-law form of return interval $T_q$ PDF. 

These results confirm that the comprehensive modeling of financial markets has to incorporate endogenous dynamics of agents as well as exogenous information or/and order flow noise. 

\section{Acknowledgments}
The author wishes to thank Dr. Aleksejus Kononovicius and Dr. Julius Ruseckas for permanent interest in this work, kind collaboration and help making some needed calculations.


\end{document}